\documentstyle[prd,aps,floats,epsfig]{revtex}
\tighten
\def\spose#1{\hbox to 0pt{#1\hss}}
\def\lta{\mathrel{\spose{\lower 3pt\hbox{$\mathchar"218$}}
     \raise 2.0pt\hbox{$\mathchar"13C$}}}
\def\gta{\mathrel{\spose{\lower 3pt\hbox{$\mathchar"218$}}
     \raise 2.0pt\hbox{$\mathchar"13E$}}}
\newcommand{\be}{\begin{equation}}
\newcommand{\en}{\end{equation}}
\newcommand{\bea}{\begin{eqnarray}}
\newcommand{\ena}{\end{eqnarray}}
\newcommand{\ex}{\mbox{e}}
\begin{document}
\draft
\twocolumn[\hsize\textwidth\columnwidth\hsize\csname 
@twocolumnfalse\endcsname
\title{Current-carrying string loop motion: \\
Limits on the classical description and shocks.}
\author{Xavier Martin$^a$ and Patrick Peter$^b$}
\address{$^a$Dpto. de F\'{\i}sica, CINVESTAV-I.P.N. A.P. 14-74-, 07000
Mexico, D.F., Mexico\\
$^b$D.A.R.C., Observatoire de Paris-Meudon, UPR 176, CNRS, 92195
Meudon, France \\ {\rm Email:~xavier@fis.cinvestav.mx,
Patrick.Peter@obspm.fr}}
\date{25 August 1999}

\maketitle

\begin{abstract} The dynamical evolution of superconducting cosmic
strings is much more complicated than that of simple Goto-Nambu
strings.  For this reason, there are only a few known analytical
solutions and no numerical ones.  The goal of this paper is to present
numerical solutions for the dynamics of planar superconducting cosmic
string loops. In most cases, a purely dynamical approach turns out to
be insufficient to describe correctly the evolution of a loop due
mainly to the appearance of shocks when spacelike currents are present
and kinks for timelike currents, leading to yet unaccounted for
quantum effects. The consequences of the quantum effects are mostly
unknown at this time because the problem requires a dynamical field
theory treatment. It is however likely that ultimately the result will
be massive radiation in the form of charge carriers from the string.
\end{abstract}

\vskip2pc]

\section{Introduction}

Cosmic strings~\cite{kibble} are topological defects~\cite{book} which
can form in the early universe during phase transitions.  The simplest
kind of cosmic string is the Goto--Nambu~\cite{GN} one which has no
internal structure and has been studied intensively in relation with
galaxy formation scenarios~\cite{network}. However, the Goto--Nambu
model is by no means the only possible one. In fact, it seems likely
that, if the Higgs boson triggering the cosmic string generating phase
transition is coupled to charged bosons or fermions, such particles
could get trapped in the strings, giving rise to conserved currents
along them~\cite{witten}. Such superconducting strings have non
trivial internal structure (because of the existence of the
current)~\cite{bps,neutral,enon0,nospring} but were thought to be
describable macroscopically by an ``elastic string'' formalism where
the internal structure is represented by just an equation of
state~\cite{carterel}. Even the first order corrections to dynamics
due to a possible electromagnetic--type current self--interaction can
be accounted for in such a way~\cite{boucleem,correcem}.  The dynamics
of such ``elastic strings'' is significantly more complicated than
that of Goto--Nambu strings, as can be expected, and for this reason
there are very few known results on them.  In particular, there is no
existing example of a simulation of the dynamics of such objects, much
less of a network of them. In fact, only the case where the chiral
limit is taken for all strings has been treated, and then very
recently~\cite{chiral}. In this paper will be shown the
first such simulation for loops and some preliminary results derived
from it.

The main known dynamical result about superconducting strings is the
existence, contrary to the case of Goto--Nambu strings, of a large
variety of dynamically stable loop equilibrium
solutions~\cite{cartin,martin}.  This has very important consequences
on the evolution of superconducting string networks since such loops
at equilibrium can accumulate and become a significant part of the
universe mass.  In fact, these loops, called ``vortons'', can put
constraints on the mass scales where superconducting strings can form,
and can also be a possible solution to the dark matter
problem~\cite{cartereta,bcdt} as well as to the ultra high energy
cosmic ray enigma~\cite{sbpp,uhecr}.

However, there are few results on the evolution of a loop toward
equilibrium.  A pioneering work was the study of the evolution of
circular loops~\cite{LA,pegabo}. It showed that, for a significant
fraction of circular initial conditions, the dynamical evolution of
the loop would drive it outside of the domain of validity of the
elastic string description to a domain where quantum instabilities
appear and likely dominate the dynamics of the string with as yet
unknown outcome~\cite{pegabo}.

In this paper, we have studied numerically the dynamical evolution of a
superconducting cosmic string loop to generalize the study conducted
in the circular case.  First, we find that in many cases, the
dynamical evolution of the loop will drive it out of the elastic regime,
thereby extending the results found in the circular case.  But even
when the loop remains in the elastic regime, it tends to develop
shocks or to fold on itself in complicated shapes, leading to a likely
non--negligible chance of spontaneous charge carrier emission by
quantum tunneling~\cite{x}. This suggests that the evolution of a
superconducting loop toward equilibrium can not be described from a
purely macroscopic point of view but must take into account purely
quantum effects which result in global charge or current loss.

These new results mean that the program will have to be improved to
include the treatment of shocks and string intercommutation. More
importantly, it begs the question of the qualitative and quantitative
consequences of the quantum effects we mentioned (which are at best
very badly known), of how they can be integrated into the existing
dynamical model, of the observability of the probable associated 
radiation, and of the consequences on the final outcome of
loop evolution.

In the following section, the equations of motion for an elastic
string, of which superconducting strings are a part, are derived.  In
the third section, these equations are reexpressed in the way most
appropriate to numerical resolution and the numerical scheme is
outlined. A fourth section discusses the various equations of state
(for Kaluza--Klein and superconducting strings) to be used afterwards
and the stability of the various perturbation modes of a string loop
endowed with such dynamics. Then, in section five, the numerical code
is applied with these various equations of states on elliptic
configurations as well as perturbed vorton states to exhibit the
probable appearance of quantum effects through either the departure of
the elastic string domain of validity, or the development of shock
waves along the string, or the existence of regions of discontinuous
curvature. Finally, we show that these effects should push loops
toward the chiral limit and conclude in the last section.

\section{The general equations  of motion}  

An elastic string is described by a two dimensional worldsheet and a
stress energy tensor which can be written as \be T^{\mu\nu}=Uu^\mu
u^\nu -Tv^\mu v^\nu ,\label{set} \en where $u^\mu$ is the timelike
unit eigenvector of the stress energy tensor, and $v^\mu$ is the other
orthogonal, spacelike, unit eigenvector.  $U$ and $T$ are the
associated eigenvalues which can be respectively identified with the
energy density and the tension of the string.  In the elastic string
model, they are related by an equation of state which describes the
substructure of the string considered, and must be derived from a
quantum field analysis~\cite{bps,neutral,enon0}. It is useful to
introduce four other thermodynamical variables, $\mu$, $\nu$, $c_{_T}^2$
and $c_{_L}^2$ which can be directly deduced from the knowledge of $U$
and $T$ by the following equations \bea
\mu & = & \frac{dU}{d\nu} , \label{munu1}\\
\mu\nu & = & U-T, \label{munu2} \\
c_{_T} & = & \sqrt{\frac{T}{U}} ,\\ c_{_L} & = &
\sqrt{-\frac{dT}{dU}}.\ena $\nu$ can be interpreted as a number
density variable, and $\mu$ as an associated effective mass variable
or chemical potential. We will see in the following that they are the
modulus of the two conserved currents in the string.  As for $c_{_T}$
and $c_{_L}$, they can be interpreted respectively as the speeds of
transverse and longitudinal perturbations along the string. They play
a fundamental role in evaluating the stability of circular loops at
equilibrium~\cite{cartin,martin}, and their being real (which means
that the corresponding perturbations are dynamically stable) defines what we
shall call the elastic regime. With the equation of state taken into
account, only one of these variables is necessary to express all six
of them.

Three other tensor fields will play an important role in the
following: the antisymmetric tangent tensor of the string worldsheet
which can be expressed using the diad of eigenvectors as \be
\epsilon^{\mu\nu} = u^\mu v^\nu - v^\mu u^\nu , \label{epsilon} \en
the first fundamental tensor, $\eta^{\mu\nu}$ which can be expressed as \be
\eta^{\mu\nu} =\epsilon^{\mu\rho}{\epsilon_\rho}^\nu = -u^\mu u^\nu
+v^\mu v^\nu ,\en and whose mixed form ${\eta^\mu}_\nu$ is the projector
along the string worldsheet, as well as the orthogonal projector on
the worldsheet \be
\perp^{\mu\nu}=g^{\mu\nu}-\eta^{\mu\nu}.\label{perp} \en

From there, the equations of motion are simply expressed as the
equation of state, which can be assumed to be solved {\it a priori} so
that only one of the thermodynamical variable is necessary to describe
them all, and the conservation of the stress energy momentum tensor
\be \overline\nabla_\mu T^{\mu\nu}=0, \en where $\overline\nabla_\mu
={\eta_\mu}^\nu \nabla_\nu$ is the covariant derivative along the
string worldsheet obtained by projecting the usual derivative on the
worldsheet. These equations of motion can be separated into an intrinsic
part which is obtained by projection along the string worldsheet, and
an extrinsic part obtained by projection perpendicularly to
it~\cite{cartin}. Note that we do not consider here the current to be
coupled with a long range interaction field such as electromagnetism.
We chose this approximation (in case it would actually be coupled with
such fields, which is not obvious in the first place) for three
reasons: first of all, it has been shown~\cite{enon0,nospring} that as
far as the internal string structure is concerned, this is utterly
negligible; then for the self interaction of a loop, it
was shown ~\cite{correcem} that its leading contribution could be
completely accounted for by means of a renormalization of the equation
of state (like the Dirac renormalization of the classical charge of an
electron). Finally, as our results show that the quantum processes are
likely to be far more important than the classical radiative
corrections concerning energy loss mechanisms, we have an
{\sl a posteriori} justification.

The intrinsic part of the system can be rewritten as the conservation
of two currents along the string, one timelike \be
\overline\nabla_\rho (\nu u^\rho)=0 ,\label{nuu} \en
and the other spacelike \be
\overline\nabla_\rho (\mu v^\rho)=0 .\label{muv} \en
One of this currents corresponds to the charged current trapped in the
string during the phase transition, the other corresponds to the
winding number of the string.  Which is which basically depends on
whether the string is in the electric sector where the charged current
is timelike, or in the magnetic sector where it is spacelike.  It is
to be remarked for future reference that these two equations can be
rewritten in adjoint form as two irrotationality equations, respectively
as \bea
\epsilon^{\rho\sigma} \nabla_\rho (\nu v_\sigma ) & = & 0, \label{nuv} \\
\epsilon^{\rho\sigma} \nabla_\rho (\mu u_\sigma ) & = & 0, \ena
where $\epsilon^{\rho\sigma}$ is the antisymmetric tangent tensor
defined in (\ref{epsilon}).

As for the extrinsic part of the system, it can be simplified using
the explicit form of the stress energy tensor (\ref{set}) as \be
{\perp^\mu}_\rho (Uu^\nu \nabla_\nu u^\rho -Tv^\nu \nabla_\nu v^\rho
)=0,\label{ortho} \en where ${\perp^\mu}_\rho$ is the projector
orthogonal to the worldsheet given by (\ref{perp}).

The system of equations thus includes the four equations (\ref{nuu}),
(\ref{muv}), (\ref{ortho}) and the equation of state.  However these
equations depend explicitly only on $u^\mu$, $v^\mu$ and the four
related thermodynamical parameters, and not on the string worldsheet
coordinates $x^\mu$ which are the unknowns of interest. In
the following section, we will explore the various ways to connect
these equations to the worldsheet coordinates through gauge fixing
conditions and find the best way to simulate the dynamics of a cosmic
string loop. 

\section{The simulation}

One possibility to solve the dynamical equations would be to choose
$u^\mu$ and $v^\mu$ as unknowns 
and integrate the worldsheet back from them, a process which however
introduces two extra unknowns. The corresponding two extra equations
come from an integrability condition~\cite{cartin} \be
{K_{[\mu\nu]}}^\rho =0 \mbox{, where } {K_{\mu\nu}}^\rho =
{\eta^\sigma}_\mu
\overline\nabla_\nu {\eta^\rho}_\sigma \en
is the second fundamental tensor of the string. 

Thus, it seems much better to choose the string worldsheet coordinates $x^\mu
(\tau ,\sigma )$ as unknown.  The timelike internal parameter of the
worldsheet $\tau$ should be identified with the time coordinate to
ensure an easy representation of the evolution of the solution in
time.  The other internal parameter $\sigma $ must be chosen so that
the expression of $u^\mu$ and $v^\mu$ as functions of the worldsheet
coordinates be straightforward. The simplest choice is to choose the
potential associated with the irrotationality equation (\ref{nuv})
which we will denote $\psi$. The meaning of this potential will depend
of the sector, electric or magnetic, considered. In the magnetic
sector, $\psi$ can be interpreted as the phase of the charge carrier
field, whereas in the electric sector it is a dual potential whose
gradient is orthogonal to that of the phase of the charge
carrier. This ensures $\psi$ to be a spacelike coordinate whatever the
sector considered. Then, there only remains three unknowns ${\bf x}(t,\psi
)$ and three equations (\ref{muv}) and (\ref{ortho}) since the fourth
(\ref{nuu}), which is equivalent to (\ref{nuv}), is automatically
solved by the choice of $\psi$ as internal parameter of the
worldsheet.

With this choice of unknowns, the other variables which appear in the
equations of motion are expressed as \bea u^\rho & = &
\frac{\dot{x}^\rho}{\dot{z}}, \\ v^\rho & = & \frac{1}{n\dot{z}}(\beta
\dot{x}^\rho +\dot{z}^2 {x'}^\rho ),\label{v}\\
\nu & = & \frac{\dot{z}}{n}, \label{nu} \ena
where dots and primes denote respectively derivative with respect to
$t$ and $\psi$, and where several new notations have been used to
contract these expressions: \bea
\dot{z} & = & \sqrt{-\dot{x}^\rho \dot{x}_\rho}, \\
z' & = &  \sqrt{{x'}^\rho {x'}_\rho} ,\\
\beta & = & {x'}^\rho \dot{x}_\rho ,\\
n & = & \sqrt{\beta^2 +{z'}^2\dot{z}^2} .\ena Note also that the
equation of state and the two equations (\ref{munu1})-(\ref{munu2})
enable to express the three other thermodynamical parameters $U$, $T$
and $\mu$ as functions of $\nu$ only and thus of ${\bf x}(t,\psi )$.
Replacing all these expressions in equation (\ref{muv}) gives \be {\bf
v}\cdot [(n^2-\beta^2 c_{_L}^2) {\bf \ddot{x}} -\dot{z}^4 c_{_L}^2 {\bf
x''}-2\beta \dot{z}^2 c_{_L}^2 {\bf \dot{x}'}]=0, \label{muvf} \en where
${\bf v}$ is just the spatial part of $v^\rho$ as defined in
(\ref{v}).

As for the extrinsic equation (\ref{ortho}), it becomes: \be {\bf
w}_{1,2} \cdot [(n^2-\beta^2 c_{_T}^2) {\bf \ddot{x}}-\dot{z}^4
c_{_T}^2 {\bf x''}-2\beta \dot{z}^2 c_{_T}^2 {\bf \dot{x}'}]=0,
\label{orthof} \en where ${\bf w}_{1,2}$ are the spatial part of two
independent quadrivectors orthogonal to the worldsheet. Two such
vectors can always be chosen among the three following vectors: \bea
w_{\perp 1}^\rho & = & (\dot{x}_1 x'_2-\dot{x}_2 x'_1,x'_2,-x'_1,0)
,\\ w_{\perp 2}^\rho & = & (\dot{x}_1 x'_3-\dot{x}_3
x'_1,x'_3,0,-x'_1) ,\\ w_{\perp 3}^\rho & = & (\dot{x}_3
x'_2-\dot{x}_2 x'_3,0,-x'_3,x'_2) .\ena 
Finally, the equation of state is assumed to be solved {\it a priori}
to get $c_{_T}^2$ and $c_{_L}^2$ as functions of $\nu$ which is itself
a function of the unknowns through (\ref{nu}).

The actual system we now have to solve can be conveniently
re--expressed in a simpler way by restricting its motion to be in a
plane, $x_3=0$ for instance. Then, only the vector $w_{\perp 1}^\rho$
gives a non trivial equation in (\ref{orthof}). The set of dynamical
equations is now given by the two equations (\ref{muvf}) and
(\ref{orthof}) which can be solved to find that the functions
$x(t,\psi)$ and $y(t,\psi)$ satisfy
\begin{eqnarray} \ddot x &=& {1\over n^2} [ A x' + B
(y'-\Delta \dot x)], \label{EOMx} \\ 
\ddot y &=& {1\over n^2} [ A y' - B (x'+\Delta
\dot y)],\label{EOMy}\end{eqnarray}
where the functions $A$ and $B$ are given by
\begin{equation} A = {\dot z^2 c_{_L}^2 \over n^2 - \beta^2 c_{_L}^2 }
[\dot z^2 (c_1 x''+ c_2 y'') +2\beta (c_1 \dot x' + c_2 \dot
y')],\end{equation}
\begin{equation} B = {\dot z^2 c_{_T}^2 \over n^2 - \beta^2 c_{_T}^2 }
[\dot z^2 (x'' y'- x'y'') +2\beta (\dot x' y' -x'\dot
y')],\end{equation}
with
\begin{equation} c_1 \equiv x'+ \Delta \dot y ,\ \ \ \ c_2 \equiv y' -
\Delta \dot x, \ \ \ \Delta \equiv \dot x y' - \dot y x'.\end{equation}

Obviously the dependence of the worldsheet position ${\bf x}(t,\psi)$
in $\psi$ must be periodic for a loop, with a constant period $L$
given by the conserved quantity associated with the conserved current
(\ref{nuu}) \be
L=\oint (\nu u^0)\sqrt{-\eta}\, d\psi=\oint d\psi.\label{paraL} \en
The string worldsheet can then be discretized on a grid with $N$
spatial points, and a variable time--step $\delta$. Since the equations of
motion are of second order, the worldsheet is then described at a
given time--step by $2N$ vectors \bea 
{\bf x}_{ij} & \simeq & {\bf x}(i\delta,\frac{j}{N}L),\\
\dot{\bf x}_{ij} & \simeq & \dot{\bf x}(i\delta,\frac{j}{N}L),\ena
and the equations of motion (\ref{EOMx})-(\ref{EOMy}), discretized
through a finite difference scheme, then enable to calculate the $2N$
vectors at the next time--step $i+1$. The time--step is automatically
adjusted by the program so that the finite difference scheme remain
stable. The total energy
\begin{equation} E_{_T} \equiv \oint d\psi {U[\nu (\psi)]\over n \dot
z^2} (n^2 -\beta^2 c_{_T}^2) \end{equation}
and angular momentum
\begin{equation} J_{_T} \equiv \oint d\psi {\beta U n \over
\dot z^2} (1-c_{_T}^2) \end{equation}
of the system, which are obviously conserved by the equations of
motion, serve as controls of the deviations of the numerical
approximation from the exact solution.  In the following, the relative
errors on the energy and angular momentum are always kept to less than
$10 ^{-6}$ over all the time evolution followed, unless specified
otherwise (as will be the case for instance due to the appearance of
uncontrollable quantum effects).

\section{Equation of state}
\label{eqs}
The equation of state is what differentiates the various elastic
string models. From a single elastic string model, there will
generally arise two different equations of state depending on whether
the current condensed inside the string is timelike or
spacelike. These two regimes of the model will be called respectively
electric and magnetic. In practice, there exists two elastic string
models of physical interest: the transsonic model (also called wiggly
Goto--Nambu model or Kaluza--Klein model), and the superconducting model.

The transsonic string is a simple model which can arise in various
ways. It is for instance the effective model to represent a wiggly
Goto--Nambu string \cite{wiggly,wiglstr}, but it can also be found to
arise from the  the 
compactification of strings in a 5-dimensional Kaluza--Klein
theory. One of the peculiarities of the wiggly Goto--Nambu model 
is that it is self--dual, meaning that the equations
of state in the two regimes are the same, given by\be
UT=m^4,\label{transs}\en
with $m$ a constant having the dimension of a mass.  Here and in what
follows, the mass $m$ is essentially the characteristic symmetry
breaking scale at which the strings form. In the case of a wiggly
Goto--Nambu string, it is also the value of the equal energy density
and tension of the bare Goto--Nambu string.

It is then possible to derive the variable $\nu$ from (\ref{munu1})
and (\ref{munu2}) as \be 
\nu = m\sqrt{U^2-m^2},\en
and from there the expressions of the two speeds of perturbations as
functions of $\nu$ as required by the equations of motion (\ref{EOMx})
and (\ref{EOMy}): \be
c_T^2=c_L^2= (1+\nu_\star^2)^{-1} ,\label{UTm4}\end{equation}
where $\nu_*=\nu/m$ is a dimensionless variable.

The fact that the two speeds of transverse and longitudinal
perturbations are equal is a characteristic of this model, and the
reason it is sometimes called the transsonic model. Another
consequence of the equality of these two speeds is that this model can
be exactly solved in flat space \cite{wiglstr}, as could be expected from the
fact that it is the effective model of a Goto--Nambu string, model
which shares this same property. The simplicity of this model makes it
ideal to test the simulation code before going on to the more
difficult superconducting string models.

The other physically interesting type of elastic string equations of
state arise when studying superconducting strings. Actual realistic
models for string superconductivity coming from field theoretical
models in four dimensions require two different mass scales, namely
the string scale $m$, already introduced, which is also the string
forming symmetry breaking scale, and the mass $m_*$ which is the one
at which the current sets up, i.e. it is roughly the mass of the
current carrier. Also such models have been found to be definitely not
self dual, and one therefore needs to know the equations of state in
the two separated electric and magnetic regimes. In this paper, we
adopt the most reasonable model of superconducting cosmic string
existing to date \cite{eqstate}.

\subsection{The magnetic regime}

In the magnetic regime, it can be seen that the equation of state
derived in~\cite{eqstate} gives \bea c_{_T}^2 & = & \frac{1}{1+\nu_*^2}
\frac{2m^2(1+\nu_*^2)^2-\nu^2 
(1-\nu_*^2)}{ 2m^2(1+\nu_*^2)+\nu^2},\\ c_{_L}^2 & = &
\frac{1-3\nu_*^2}{1+\nu_*^2} ,\ena where now $\nu_*=\nu /m_*$. Here,
$\nu^2_*<1/3$ to insure that $c_{_L}^2$ remains everywhere positive.
It is dynamically possible for $\nu$ to go above this value (as will
be seen below), but then the string becomes unstable with respect to
longitudinal perturbations and the elastic string description we
adopted to describe the dynamics of the string breaks down due to
quantum instabilities in the string. As we will see below, most of the
string loop trajectories, independently of the initial conditions,
develop shock waves, the description of which is beyond the scope of
the simulation presented here.

\subsection{The electric regime}

In the electric regime, the best fit to the equation of state which
also satisfies the phase frequency threshold
behavior~\cite{neutral,wall} was also derived in~\cite{eqstate} and
gives for our purpose \bea c_{_T}^2 & = & 1-\frac{2X_*}{[2m^2/m_*^2+\ln
(1-X_*)](1-X_*)+2X_*},\\ c_{_L}^2 & = &
\frac{1-X_*}{1+X_*} \\ X_* & = &
\frac{2\nu_*^2+1-\sqrt{4\nu_*^2+1}}{2\nu_*^2},\ena where, as
previously, $\nu_* =\nu /m_*$, and $\nu$ is bounded $\nu <
\sqrt{\ex^{2m^2 /m_*^2}-1}$, but this time
so that $c_{_T}^2 >0$. Again, it is dynamically possible
for $\nu$ to go above this value (as will be seen below), but then the
string becomes unstable with respect to transverse perturbations and
the elastic string description we adopted to describe the dynamics of
the string breaks down due to quantum instabilities in the string. In
this regime, near the critical current (which can become large with
respect to $m$ contrary to the magnetic regime), if the current in the
string can interact electromagnetically, the dynamical corrections due
to the current self--interaction become large and an effective equation
of state must be used to take them into account to first order
\cite{correcem,pegabo}. This effective equation of state is not
expected to change significantly the qualitative results of this paper
and will therefore be ignored in the following.

\subsection{Stability}

\begin{figure}
\centering
\epsfig{figure=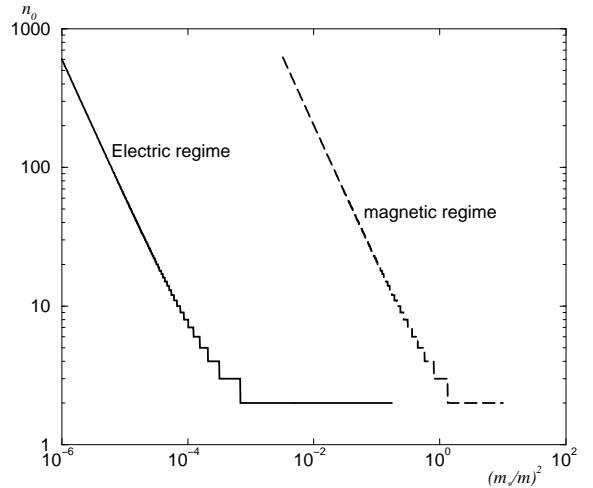, width=9cm}
\caption{First unstable equatorial mode number found when the current
grows up from $0$, for the electric (full line) and magnetic (dashed line)
regimes, as a function of the square mass ratio $(m_\star /m)^2$.
\label{stabN}}
\end{figure}

Circular rotating configurations are known to be equilibrium states,
often called vortons, with stability depending on the perturbation
mode number and the two perturbation velocities $c_{_L}$ and
$c_{_T}$. In other words, whether a mode is stable or not essentially
depends on the relevant equation of state. Note that, although only
equatorial perturbations (i.e. in the plane of the loop) are
considered here, transverse perturbations (i.e. orthogonal to the
plane of the loop) are always stable and would not therefore introduce
any new special behavior.  For the Kaluza--Klein self--dual model
discussed above, stability is ensured whatever the mode number since
in this case $c_{_L}=c_{_T}$. This fact will enable us to describe any
perturbed equilibrium loop configuration without difficulty, the loop
being unable to leave the region where the elastic description is
valid. However, this is no longer the case for realistic
superconducting string models, such as the one described above, since
the squared velocities $c^2_{_L}$ and $c^2_{_T}$ can now become
negative, in which case the macroscopic description becomes
insufficient and microscopic (quantum) effects should therefore be
taken into account. Moreover, the perturbation velocities are no
longer always equal, and one must use the full stability
formalism~\cite{martin} to derive the complicated relation between the
mode numbers $n$, corresponding to perturbations of the type
\be r=R_0+\delta R \exp [i (\omega t -n\theta )],\label{equatorial} \en
with $(r,\theta)$ the polar coordinates in the loop plane and $\omega$
the angular velocity of the mode, and their stability. As both these
models basically depend on a single parameter, namely the mass ratio
$m_\star/m$, we present here the regions for which instability is
likely to occur as a function of the mass ratio.

Fig.~\ref{stabN} shows the first unstable mode number found when the
current grows from $0$, for each model,
as a function of the mass ratio $(m_\star/m)^2$. The curve in the
electric regime stops at $m_*^2/m^2\simeq 0.65$ because beyond this
point all vortons are stable \cite{xm}, and the first unstable mode therefore
does not exist. It can also be 
argued that for moderately small ratio $m_\star \lta 10^{-3} m$, the
first unstable modes are found for $n\sim 10^3$, and these modes 
require large amounts of energy to be excited. This means that if, as
proposed by many realistic superconducting string models, the current
carrier is reasonably less massive than the string forming Higgs
field, then the vortons formed are likely to be almost stable, even
from a cosmological point of view. We have therefore used the results
of Figs.~\ref{stabN} to investigate only those low $n$ (in practice
less than ten) that are easily excited and unstable. This means we
have concentrated our attention in what follows on square mass ratios
ranging from a $10^{-4}$ to unity. This does not mean that this
parameter should be found in such a range but simply that this is the
region of parameter space where interesting effects are supposed to
take place.

A final conclusion that can be drawn from Fig.~\ref{stabN} is that the
instability regions are widely separated between spacelike and
timelike currents. This mean in practice that given the effective
value of the mass ratio, the evolution of the spacelike and timelike
distribution could be qualitatively very different, as for $m_\star /m
\lta 10^{-2}$, the magnetic vortons are almost stable, with first
unstable mode greater than a thousand, while the electric vortons
suffer instabilities already for the $n=2$ mode, although it should be
remarked that those vortons carrying very strong timelike currents are
expected to be completly stable~\cite{xm,xp}

\section{Evolution}

The code was first tested on some of the few known exact solutions.
For instance, equilibrium states are left stationary for as long as
we care to run the program. Similarly, stable circular modes are
found stable by the program while unstable ones do start leaving
equilibrium almost immediately. The program is not capable to
determine with certainty what is the long range evolution of these
unstable modes due to the fact, discussed below, that other effects
should be taken into account in their description.  Another clue that
the program is performing well is that it conserves circularity for as
long as we care to run it and reproduces exactly the results of
Ref.~\cite{pegabo}.

Since circular solutions have been analyzed analytically in
detail~\cite{pegabo}, we show in this chapter the numerical evolution
of elliptic initial configurations and of various perturbed vorton
states, stable and unstable.

First was investigated the fate of elliptic loops and equatorial
perturbations of vortons (\ref{equatorial}) with the transsonic
equation of state, as it is simpler and yields only stable vorton
modes. Elliptic configurations present the advantage of showing all
the non--linear effects right from the beginning of the simulation,
whereas the stable equatorial modes enable to check the program and
give a reference before studying unstable modes with the
superconducting equations of state.

Then the more complicated superconducting equations of state presented
in chapter \ref{eqs} are investigated, again using elliptic
configurations and equatorial perturbations of vortons. For elliptic
initial configurations, in both regimes (magnetic or electric),
examples of time evolution similar to that of the transsonic string
were obtained, and are not shown again. However, in some instances,
completely different (and in some cases unexpected) dynamics were
found, leading to drastic modifications in our understanding of the
dynamics of a cosmic string loop. As for perturbed vortons, since
stable ones obviously yield the same results as in the transsonic
case, only unstable perturbation modes are shown. The latter start
evolving in the linear approximation, but because they are unstable
must grow exponentially, becoming more and more non--linear as time
passes. A systematic study was done to find the generic behavior of
these unstable modes.

\subsection{Kaluza--Klein strings}

\begin{figure}
\epsfig{figure=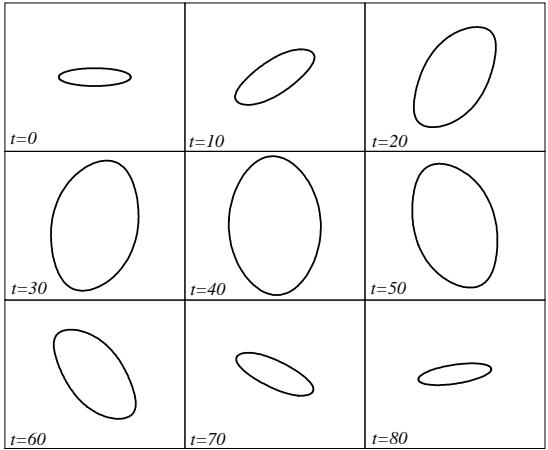, width=9cm}
\caption{Elliptic configuration evolution for a
Kaluza--Klein equation of state. Here, $e=0.3$ and $\nu^2_{\star 0}
=1$. The periodic evolution reveals that the loop size increases while
its shape is conserved.\label{conf1}}
\end{figure}

\begin{figure}
\epsfig{figure=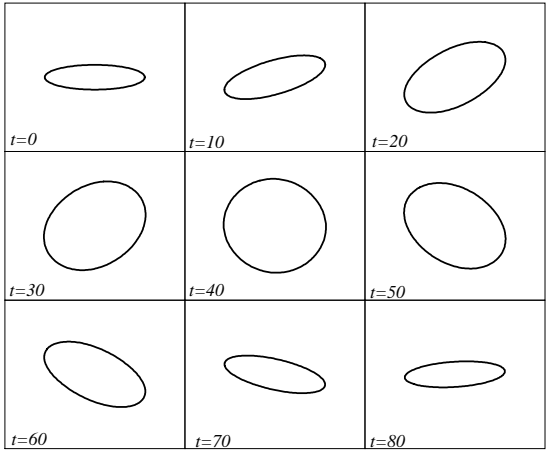, width=9cm}
\caption{Same as Fig.~\ref{conf1} with now $\nu^2_{\star 0}=0.1$. In
this particular case, the shape also changes, becoming 
almost circular at about half the period.\label{conf2}}
\end{figure}

The first series of simulations performed concern
elliptic loops, for which the initial configuration reads
\begin{equation} x(t=0,\psi) = R \cos \left( {\psi\over R\gamma
\nu_0}\right),\label{xell}\end{equation}
\begin{equation} \dot x(t=0,\psi) = -{c_{_T} \over 1+\varepsilon} \sin
\left( {\psi\over R\gamma \nu_0}\right),\label{xdotell}\end{equation}
\begin{equation} y(t=0,\psi) = R \, e \sin \left( {\psi\over R\gamma
\nu_0}\right),\label{yell}\end{equation}
\begin{equation} \dot y(t=0,\psi) = {c_{_T} \over 1+\varepsilon} \cos
\left( {\psi\over R\gamma \nu_0}\right),\label{ydotell}\end{equation}
where $\nu_0$ is the initial value of the state parameter, assumed
constant along the worldsheet, $e$ is the ellipticity of the
configuration (for $e=1$, one gets a circle of radius $R$), the
spacelike coordinate $\psi$ varies from zero to $2\pi R\gamma \nu_0$, and
$\gamma=(1-\dot x^2 -\dot y^2)^{-1/2}$ is the (constant) initial
Lorentz factor; the dimensionless parameter $\varepsilon$ measures the
deviation in velocities from the equilibrium state obtained when
$e=1$ and $\varepsilon=0$. 

The first two figures, namely Figs.~\ref{conf1} and \ref{conf2}, show
two different kinds of such equilibrium configurations for which
$\varepsilon =0$. These figures differ only in their initial state
parameter $\nu_0$ and are presented to show that the ellipticity can
be, either conserved as in Fig.~\ref{conf1}, or made to vary greatly
as in Fig.~\ref{conf2} where the loop becomes almost circular halfway
through its periodic evolution.

\begin{figure}
\centering
\epsfig{figure=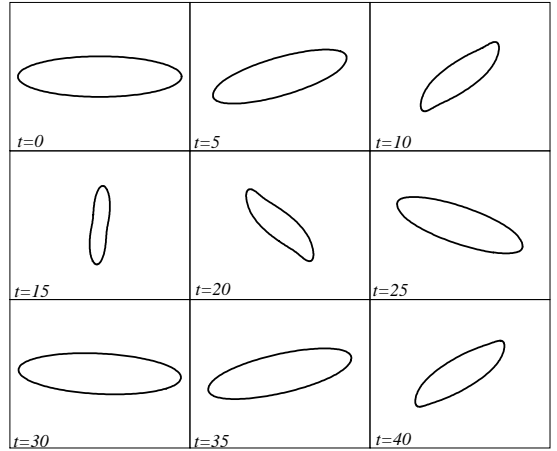, width=9cm}
\caption{Same elliptic configuration evolution as on Fig.~\ref{conf1}
but with the velocity parameter set to $\varepsilon =0.1$. It is seen
that the shape is modified drastically even though the departure from
equilibrium is relatively modest.\label{conf3}}
\end{figure}

\begin{figure}
\centering
\epsfig{figure=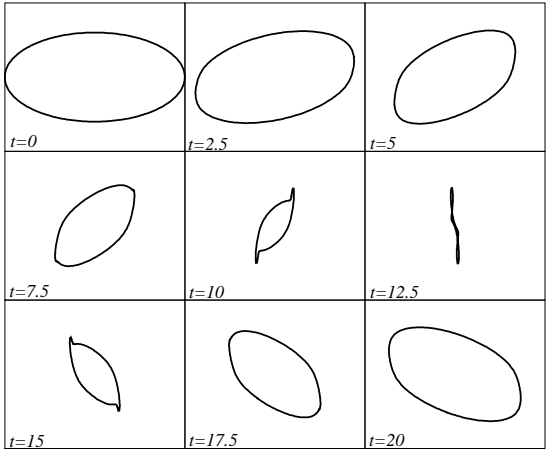, width=9cm}
\caption{Same as Fig.~\ref{conf3} but with a larger velocity deviation
$\varepsilon=1$. The loop 
contracts so much as to attain infinite curvature, i.e. points where
intercommutation should be taking place.\label{conf4}}
\end{figure}

Getting further away from equilibrium is very simply achieved
by increasing the value of the parameter $\varepsilon$. On
Fig.~\ref{conf3} is shown the evolution of a loop with the same
initial conditions as on Fig.~\ref{conf1} but with $\varepsilon
=0.1$. During the loop evolution now, as the initial velocity has been
slightly reduced, the loop size decreases, which was expected. Shapes
similar to those of Figs.~\ref{conf1} and \ref{conf2} are produced,
with a loop size increasing, for negative values of
$\varepsilon$. What is more interesting however is the shape of the
loop which is very strongly deformed during its quasi--periodic
motion. As is seen on Fig.~\ref{conf3}, regions of relatively
large curvature tend to be formed. These points are good candidates
for charge carrier emission.

\begin{figure}
\centering
\epsfig{figure=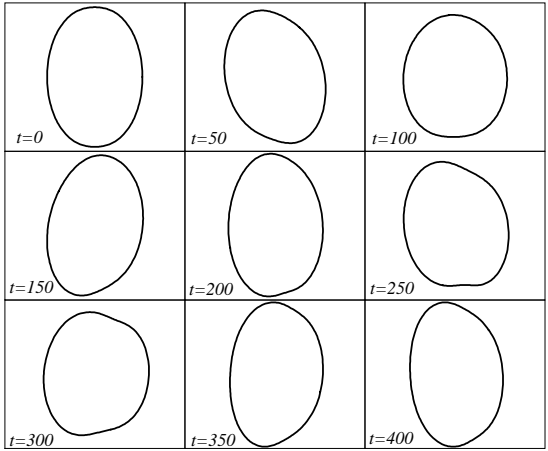, width=9cm}
\caption{$n=2$ perturbation mode of a transsonic vorton with a rather
high amplitude of $\delta R/R=0.6$.\label{conf5}} 
\end{figure}

Exaggerating even more the value of $\varepsilon$ yields
Fig.~\ref{conf4}, where cusps are now formed, i.e. points where the
curvature diverges, and the string loop shows points where
intercommutation should be occuring. The treatment of the ensuing
evolution is beyond the scope of the elastic string description and
therefore, even though the simulation goes on very well at these
points, we believe the relevant physics not to be taken into account
properly from this time on. For instance, charge carriers will also be
emitted at such points. In the case of the wiggly Goto--Nambu string
model, this emission will take the form of microscopic loops produced
by intercommutation of the microstructure moving in the string. As
will be discussed further down, this emission should be evaluated as
it is a fairly generic situation in this model but also in the
superconducting string one.

\begin{figure}
\centering
\epsfig{figure=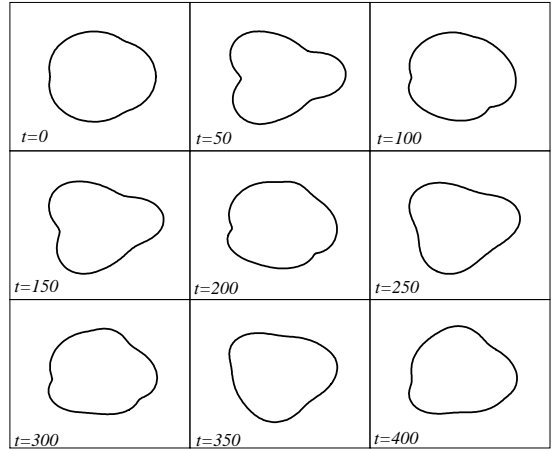, width=9cm}
\caption{Same as Fig.~\ref{conf5} for the mode $n=3$.\label{conf6}}
\end{figure}

\begin{figure}
\centering
\epsfig{figure=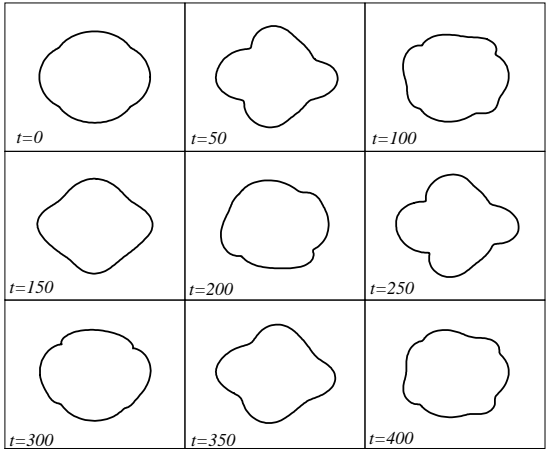, width=9cm}
\caption{Same as Fig.~\ref{conf5} for the mode $n=4$ and $\delta R/R$
lowered to $0.45$.\label{conf7}}
\end{figure}

Let us now turn to the fate of circular ``vorton'' states with
equatorial perturbations of the form
(\ref{equatorial}). Figs.~\ref{conf5}, \ref{conf6} and \ref{conf7}
show the evolution for the modes $n=2$, $3$ and $4$. Since we have
seen that transsonic vortons are always stable, a
high perturbation value for $\delta R/R\simeq 0.5$ could be used. Here
again, it is found that the motion is quasi--periodic and that the shape is
very strongly modified during a period. It is interesting to note
however that even though the evolution we investigated involved non
linear effects because of the large amplitude used, the modes remain
almost completely decoupled throughout the loop trajectory. This is to be
contrasted to what happens with the other equation of state
corresponding to actual particle condensates in strings and to which
we now turn.

\subsection{The magnetic shock}

\begin{figure}
\centering
\epsfig{figure=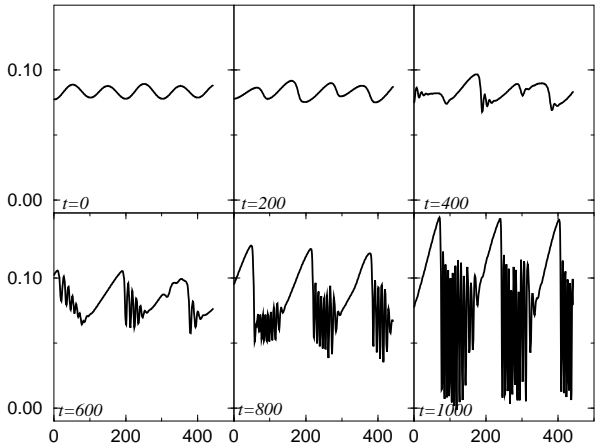, width=9cm}
\caption{Time evolution of the state parameter $\nu^2_\star$
along the string worldsheet for an unstable mode $n=3$ in the magnetic
regime. This figure is for the magnetic equation of state with mass
ratio $m_\star/m = 1$ and initial mean state parameter $\nu_\star
=0.3$. It shows the development of a physical shock as it is clear that
$\nu$ is indeed a gauge invariant quantity. When the discontinuity
becomes too important to handle for the program, the Gibbs phenomenon
occurs, as expected, and the simulation ends.
\label{choc1}}
\end{figure}

Let us start with an unstable mode $n=3$ as exemplified on
Fig.~\ref{choc1}. On this figure is plotted not the actual string loop
trajectory, as nothing visible to the naked eye could then be seen:
even the amplitude of the perturbation does not have time enough to
increase before the important effect takes place, and what is seen is
just the ordinary rotation of the unperturbed loop.  Instead, what we
show is the gauge invariant and rescaled state parameter $\nu_\star^2$
as a function of the spacelike string internal coordinate $\psi$, in
our case the point label. Although this coordinate is, by its very
definition, not gauge invariant, the same results apply when
$\nu_\star$ is plotted again say, the $x$ position coordinate of the
loop, with the advantage of showing an unwrapped picture. This
simulation used $500$ space points and clearly shows that a shock wave
is being formed at three different points (whose precise position is
physically irrelevant but that are located at equal distances from
each other, thereby preserving the original symmetry) along the
string worldsheet. This is a generic conclusion for unstable modes: a
mode of index $n$ will develop, in the magnetic regime, $n$ shock
fronts. As can be seen from the figure, when the discontinuity forms,
small oscillations appear which are not physical but instead reflect
the occurrence of the well--known Gibbs phenomenon, produced by the
simulation scheme which cannot, with a uniform grid, describe
shocks. This figure also shows that the shocks tend to increase the
amplitude of variation of the current.

\begin{figure}
\centering
\epsfig{figure=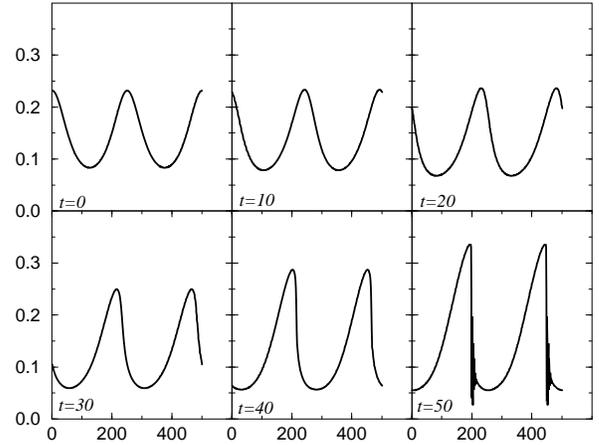, width=9cm}
\caption{Same as Fig.~\ref{choc1}, but for an elliptic initial
configuration with $e=0.6$. Again, shocks develop, and in much
the same way as if it was an unstable $n=2$ initial configuration.
\label{choc2}}
\end{figure}

Fig.~\ref{choc2} shows a similar evolution pattern for an initially
elliptic configuration. In this particular case, it is clear from the
periodicity that the leading mode would be $n=2$, even though all the
unstable modes are actually present in this initial configuration.
This is translated in the figure by the presence of only two shock
fronts. In this figure, and contrary to the previous case, the shock
can actually drive the configuration outside the range of validity of
the elastic string description, as $\nu_\star^2$ can reach values
greater than a third for which the longitudinal perturbation velocity
ceases to be real. Two possibilities are thus opened: either the shock
occurs while the string is still in the elastic regime, or the
longitudinal dynamical instability occurs before the shock is
completed. In both situations however, we believe our simulation to
stop giving an accurate description of the phenomena that should be
taking place.

To continue further the simulation would require a semi--classical
treatment of the quantum effects that are expected to take place where
the shocks form. Most probably, the leading effect will be to emit one
current--carrier particle anywhere the state parameter becomes
discontinuous. This fact entitles us to cast serious doubts as to the
actual formation of shocks in cosmic string loops: contrary to its
analog in fluid dynamics, a string shock lives in two dimensions, the
string worldsheet, that is embedded in the four dimensional space
time, thus opening the phase space to particle emission. As a result,
the energy contained in the shocks can be released in such a way as to
smooth it out. From subsection \ref{chirall}, such an effect would lead to
current loss, so that the loop distribution must favor the chiral limit in
network simulations. More detailed studies must however be carried on
before any definite conclusion can be reached on that point.

\subsection{The electric kink}

\begin{figure}
\centering
\epsfig{figure=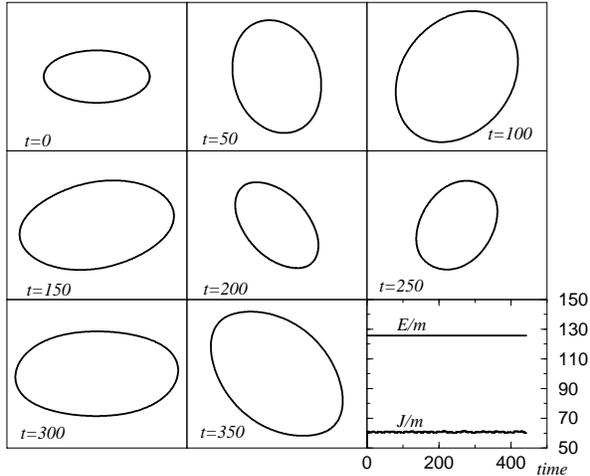, width=9cm}
\caption{Time evolution of essentially the same elliptic configuration
as on Fig.~\ref{choc2} but in the electric regime instead. In this
regime, no shock can develop, 
as shown in Fig.~\ref{elec1}. The bottom--right corner shows the total
energy and angular momentum of the loop versus time, and they are
almost exactly conserved during the quasi--periodic evolution.
\label{elec1XY}}
\end{figure}

\begin{figure}
\centering
\epsfig{figure=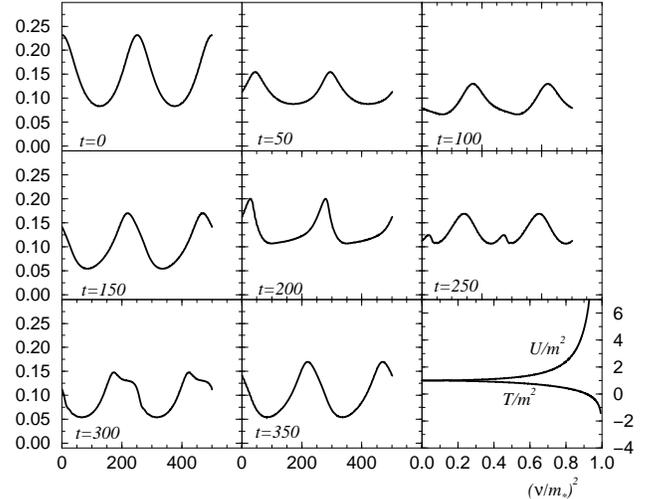, width=9cm}
\caption{Time evolution of the state parameter $\nu_\star^2$ against
the point labeling function $\psi$ for Fig.~\ref{elec1XY}. Contrary to
what happened in the magnetic case, the harmonic developped by the
state parameter is smoothed out. The bottom--right corner recalls the
equation of state $U/m^2$ and $T/m^2$ as functions of $\nu_\star^2$.
\label{elec1}}
\end{figure}

To see what happens in the electric regime and check the result of
Ref.~\cite{vlasov} according to which no shock should take place, we
have performed a simulation with exactly the same initial parameters
as Fig.~\ref{choc2}, but now for a timelike current and the
equation of state of the electric regime. Fig.~\ref{elec1XY} shows
the spatial configuration of the loop, with an evolution very similar
to that of the transsonic string loop, while Fig.~\ref{elec1} shows
the evolution of the state parameter. This completely stable and quasi
periodic motion develops an harmonic frequency which is subsequently
smeared out. This effect, which prevents the apparition of shocks, was
found to occur any time a shock was expected to form in the electric
regime. As a result, we claim that, indeed, no shock can be formed in
a timelike current carrying cosmic string loop, at least through
ordinary dynamical effects.

\begin{figure}
\centering
\epsfig{figure=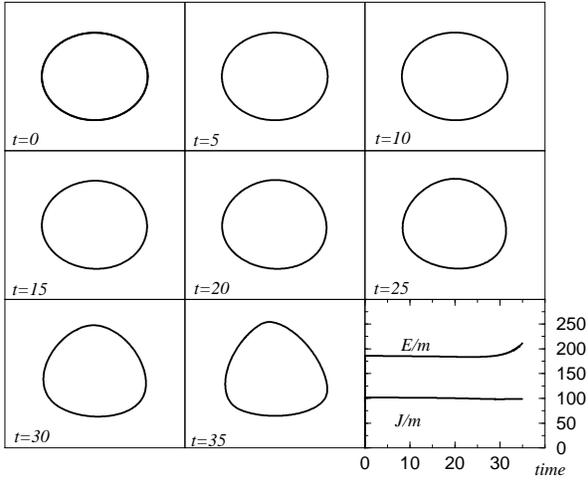, width=9cm}
\caption{Time evolution of the $n=2$ unstable mode for an electric loop
with mass ratio $(m_\star/m)^2 =0.1$ and initial value of the state
parameter $\nu^2_{\star 0} =0.3$. The loop shape is modified until
three kinks, i.e. regions of discontinuous curvature, are formed and the
total energy is no longer conserved, meaning the elastic string
description is insufficient.
\label{elec2XY}}
\end{figure}

\begin{figure}
\centering
\epsfig{figure=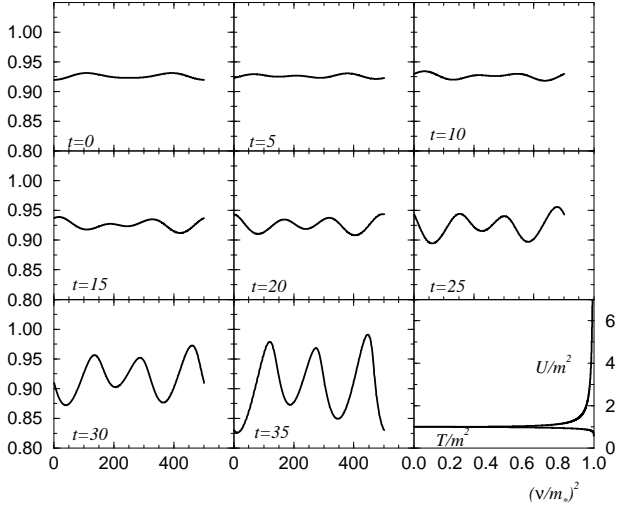, width=9cm}
\caption{Same as Fig.~\ref{elec2XY} for the state parameter versus the
point labeling. As $\nu_\star$ goes to large values, approaching
unity, the dynamics drive the tension to negative values.
\label{elec2}}
\end{figure}

\begin{figure}
\centering
\epsfig{figure=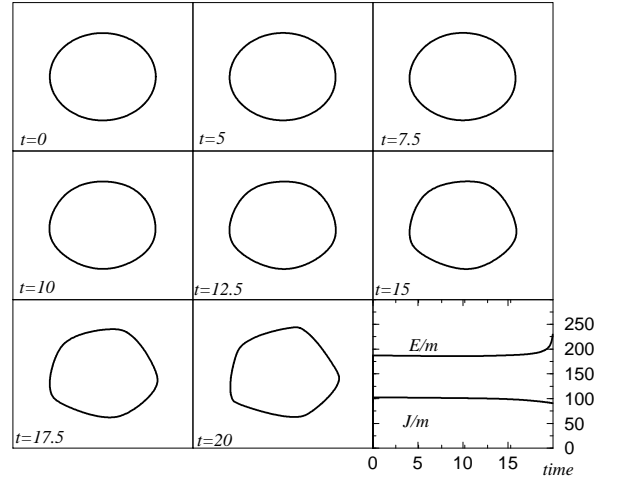, width=9cm}
\caption{Same as Fig.~\ref{elec2XY} for a mode $n=3$. In this case,
one finds that five kinks develop.
\label{elec3XY}}
\end{figure}

\begin{figure}
\centering
\epsfig{figure=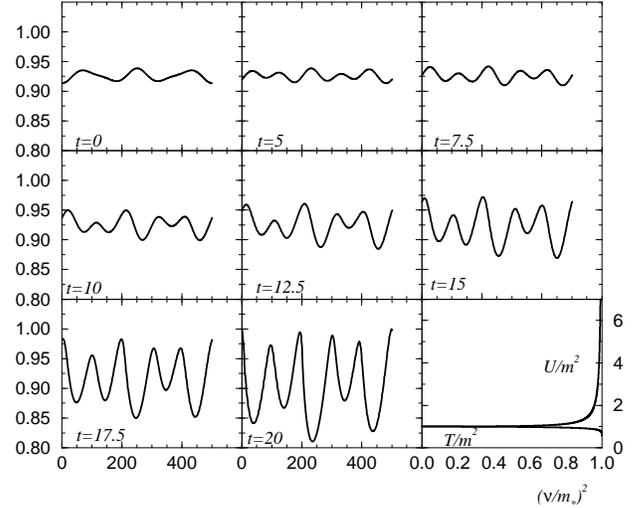, width=9cm}
\caption{Same as Fig.~\ref{elec3XY} for the state parameter versus the
point labeling.
\label{elec3}}
\end{figure}

Instead of shocks, electric string loop evolution exhibits an
unexpected feature, namely the appearance of what is usually called
kinks. These are regions, shown for instance on Figs.~\ref{elec2XY}
and \ref{elec3XY}, where the curvature of the string becomes
discontinuous, and the state parameter tends to unity
(Figs.~\ref{elec2} and \ref{elec3}), limit where, as recalled on these
latter figures, the tension reaches negative values. Thus, at these
points the string leaves the domain of validity of our dynamical
formalism likely resulting, again, in particle emission and as shown
in subsection \ref{chirall}, evolution of the end vorton toward the
chiral limit and thus better stability.

An astonishing point is the number of kinks formed: we have found
generically that an unstable equatorial mode of index $n$ yields
exactly $2n-1$ kinks. We could not find any explanation for this
completely general result.

Therefore, it appears to be unavoidable that, in both domains, cosmic
string loops will radiate charge carriers.  The exact processes
whereby it can do so has to be explored in more details and in
particular quantified.  As long as this has not been done, the
consequences on the evolution and decay of superconducting cosmic
string loops will remain sketchy at best.

\subsection{The chiral limit}
\label{chirall}
As shown in (\ref{paraL}), the space--like coordinate $\psi$ can be used to
describe the conserved charge associated with the conserved current
$\nu u^\rho$. For a spacelike current, it is the conserved winding
number $N$ of a string loop such that~\cite{pegabo}
\begin{equation} N\propto \oint d\psi,\end{equation}
where the proportionality factor depends on the string length. For a
timelike current, the same integral would yield a result similarly
proportional to the conserved particle number $Z$ present on the
string. However, the physically relevant quantity is instead the
invariant number
\begin{equation} I^2 = N^2 - Z^2, \end{equation}
whose sign determines whether the current is spacelike (positive sign)
or timelike (negative sign). In the gauge and reference frame chosen
here, one of these parameters is set to vanish ($Z$ for the magnetic
regime and $N$ for the electric one), while the other can be obtained
by integrating the state parameter over the worldsheet.

In the magnetic regime, when quantum effects are taking place where
shocks form, they lead to the emission of some part of the current,
with the result of lowering the winding number $N$. But as in this
case $Z=0$, one sees that the invariant quantity $I^2$ is lowered in
the process. If the resulting loop were still unstable, it would
repeat this process, and one can conjecture that the resulting stable
vorton configuration will be close to chiral, for which $N^2 = Z^2=0$.

In the electric regime, the same apply with now $N$ set originally to
zero and $Z$ decreasing as the string emits particles at each
kink. Therefore, again, one find that $|I^2|$ is lowered, up to the
point where the chiral limit could be reached from the opposite end. 

\section{Conclusions}

We have explored the full nonlinear evolution of superconducting
cosmic string loops with rational and logarithmic equations of
state~\cite{eqstate}. Starting with elliptic configurations, i.e. far
from equilibrium, and varying the state parameter in the
initial conditions allowed us to examine many loop trajectories. We
have also investigated the fate of unstable equatorial modes.

In the magnetic regime where the tension is constrained to be
positive~\cite{nospring}, some string loops were found to exhibit
shocks, seen as discontinuities in the state
parameter. Near these shocks, the longitudinal perturbation
velocity may become imaginary, implying instabilities at the
classical level, to be presumably understood later in terms of
massive radiation, i.e., quantum instabilities leading to charge
carrier emission.

For electric strings for which the constraint is $c_{_{L}}^2>0$, some
configurations 
were found to evolve to the point where the tension becomes negative, so
that transverse instabilities would ultimately be transformed into,
again, massive radiation.

Finally, all the configurations that were not leading to the
previously described effects, i.e., those for which the classical
analysis was found valid throughout their evolution, were shown to
give rise to regions of very high curvature radius, a situation known
to enhance considerably the escape probability for the trapped charge
carrier particles.

The purely classical treatment of the evolution of current--carrying
cosmic strings has therefore been proved insufficient. Further studies
are now needed to find some ways of incorporating the quantum effects
whose significance has been emphasized here. Once this is achieved,
the evolution scheme presented here might be used in order to evaluate
the rate of massive radiation during the lifetime of a characteristic
loop, and its fate when back reaction is incorporated.  With that
knowledge, it might in turn become possible to seriously estimate the
spectrum of this massive radiation. Eventually, these results should
enable to derive constraints on the parameters of the problem by
comparison with observed cosmological datas. Note that, for the time
being, it is not clear whether the effects exhibited here have a
tendency to enhance or lower the vorton formation rate. On the one
hand, if radiation is very efficient, it could be argued that many
configurations might turn out to decay into massive radiation and
curentless Goto--Nambu
strings. On the other hand, if radiation is not that efficient, it
could in fact accelerate the vorton formation rate by helping to form
the final vorton state more quickly, and by driving this final vorton
state toward the chiral limit where it is stable. The resulting
distribution would then be very close to the chiral limit.

\end{document}